\documentclass[11pt]{article}
\pdfoutput=1
\usepackage[utf8]{inputenc}
\usepackage{multirow}
\usepackage{amsmath, amsfonts, amssymb}
\usepackage{comment}
\usepackage{graphicx}
\usepackage{psfrag}
\usepackage{amsthm}
\usepackage[usenames,dvipsnames,svgnames,table]{xcolor}
\usepackage{enumerate}
\usepackage{arydshln}
\usepackage{soul}
 \usepackage{slashed}
\usepackage{subfig}
\usepackage{mdframed}
\usepackage{mathrsfs}
 \usepackage{a4wide}
  \usepackage{tikz}
  \usepackage{tikz-cd}
  \usetikzlibrary{shapes.geometric}
  \usepackage{tcolorbox}

  \usepackage{color}
  \definecolor{dark-gray}{gray}{0.20}
  \definecolor{gray}{gray}{0.30}
  \definecolor{light-gray}{gray}{0.80}
  \definecolor{dark-red}{rgb}{0.7,0,0}
  \definecolor{dark-green}{rgb}{0.1,0.4,0}
  \definecolor{dark-blue}{rgb}{0.3,0.3,0.7}
  \definecolor{light-blue}{rgb}{0.8,0.8,1}
      \definecolor{swamp}{RGB}{240, 199, 197}

  \usepackage{pifont}
\usepackage{setspace}

\mdfdefinestyle{statementstyle}{
   frametitlerule = true,
   frametitlefont = {\normalfont\bfseries},
   frametitlebackgroundcolor = blue!10,
}
\mdtheorem[style=statementstyle]{statement}{Finiteness Conjecture}

\numberwithin{equation}{section}

\newcommand{\be}{\begin{equation}}
\newcommand{\ee}{\end{equation}}
\newcommand{\eq}[1]{(\ref{#1})}

\def\be{\begin{equation}}
\def\ee{\end{equation}}
\def\bea{\begin{eqnarray}}
\def\eea{\end{eqnarray}}

\newcommand{\beq}{\begin{equation}}  \newcommand{\eeq}{\end{equation}}
\newcommand{\bal}{\begin{aligned}}   \newcommand{\eal}{\end{aligned}}
\def\beqa{\begin{eqnarray}}
\def\eeqa{\end{eqnarray}}

\def\simleq{\; \raise0.3ex\hbox{$<$\kern-0.75em
      \raise-1.1ex\hbox{$\sim$}}\; }
   \def\simgeq{\; \raise0.3ex\hbox{$>$\kern-0.75em
      \raise-1.1ex\hbox{$\sim$}}\; }

\numberwithin{equation}{section}

\usepackage{jheppub}
\usepackage{hyperref}
\usepackage{cleveref}

\hypersetup{
	colorlinks=true,
	linkcolor=dark-blue,
	citecolor=dark-red,
	urlcolor=dark-green,
	linktoc=page
}

\theoremstyle{remark}

\crefname{appendix}{Appendix}{Appendices}

\title{\centering Finiteness and the Swampland}

\author{Yuta Hamada$^1$,} \author{Miguel Montero$^1$,} 
\author{Cumrun Vafa$^1$,}
\author{and Irene Valenzuela$^{1,2}$} \affiliation{$^1$Department of Physics, Harvard University, Cambridge, MA 02138, USA}
\affiliation{$^2$Instituto de F\'{i}sica Te\'{o}rica UAM-CSIC and Departamento de F\'{i}sica Te\'{o}rica, Universidad Aut\'{o}noma de Madrid, Cantoblanco, 28049 Madrid, Spain }

\abstract{We view and provide further evidence for a number of Swampland criteria, including the Weak Gravity Conjecture, Distance Conjecture and bounds on the finiteness of the quantum gravity vacua from the prism of the finiteness of black hole entropy.  Furthermore we propose that at least all of these Swampland statements may be more fundamentally a consequence of the finiteness of quantum gravity amplitudes.}

\begin{document}
\hypersetup{pageanchor=false}
\makeatletter
\let\old@fpheader\@fpheader

\makeatother

\maketitle

\newcommand{\remove}[1]{\textcolor{red}{\sout{#1}}}

\newpage

\section{Introduction}

In going from consistent quantum theories without gravity to ones including gravity, various restrictions arise, known as Swampland constraints (see \cite{Brennan:2017rbf,Palti:2019pca,vanBeest:2021lhn,Grana:2021zvf} for reviews).  For example, the Weak Gravity Conjecture (WGC) \cite{Arkani-Hamed:2006emk} suggests that the mass of charged states cannot be arbitrarily large.  Or the Distance Conjecture \cite{Ooguri:2006in} suggests that the length in field space for a consistent EFT, which can be infinite without gravity, becomes finite in the presence of gravity. Similarly, the number of degrees of freedom of a quantum field theory, which can be arbitrarily large without gravity, is bounded if we include gravitational effects.  Related to this, with a suitable definition of counting of theories, we expect to have only a finite number of consistent quantum theories of gravity (actually, only one if assuming the Cobordism Conjecture \cite{McNamara:2019rup}) instead of infinitely many as is the case without gravity.  Thus, many (and perhaps all) of the Swampland conditions somehow emerge from a suitable replacement of infinity by a finite number.   This of course is technically achieved through replacing $M_{p}$ from infinity by a finite value.  The aim of this paper is to attempt to initiate a program to view all the Swampland criteria from the prism of finiteness and also to offer an explanation for this, thus taking a step in unifying the Swampland conjectures. 

Black holes and their thermodynamical properties have played a central role in motivating many of the Swampland conjectures.
For example the non-vanishing of black hole entropy has been used to motivate completeness of gauge charge spectrum.  
Most recently, the fact that the entropy of a black hole is not infinite has been used \cite{Hamada:2021bbz,Bedroya:2021fbu} to argue that the moduli space of $p$-brane probes in a quantum theory of gravity has a finite diameter for $p< d-2$.   If we view scalars in the bulk gravity as the moduli of a ``$(d-1)$-brane probe" this would have led to the Distance Conjecture if we could extend the domain of validity of the argument in \cite{Hamada:2021bbz,Bedroya:2021fbu} to higher values of $p$. However it turns out that this is not as straightforward.  Nevertheless in Section \ref{sec:blackhole} of this paper we provide such a link between the finiteness of black hole entropy with the Distance and Weak Gravity Conjectures. We show that the EFT cutoff must decrease as dictated by the conjectures to avoid a violation of entropy bounds coming from small black holes.   In Section \ref{sec:emergence} we review the Emergence proposal for the Distance Conjecture \cite{Grimm:2018ohb,Heidenreich:2018kpg} which is very much in the same spirit as demanding finiteness of the diameter of the field space in the UV theory.  We also relate this to the asymptotic behavior of potentials.
In Section \ref{sec:vacua} we discuss these finiteness ideas with the finiteness of the number of quantum gravity vacua, carefully defining what we mean by finiteness.  Finally in Section \ref{sec:finiteness} we conclude by suggesting that these finiteness features are related to the finiteness of quantum gravity amplitudes.

\label{sec:intro}

\section{Distance and Weak Gravity Conjectures from the finiteness of black hole entropy}  \label{sec:blackhole}

In this Section we present a bottom-up argument for the Distance Conjecture based on black hole physics and the Bekenstein bound, that applies to infinite distance limits in which a gauge coupling goes to zero. It also serves as an argument for a Tower version of the WGC \cite{Heidenreich:2015nta,Heidenreich:2016aqi,Andriolo:2018lvp} (and its magnetic version) which does not rely on stability of black holes but on entropic considerations. This is, to our knowledge, the first time that entropy bounds are used to argue for the strong versions of these Swampland conjectures\footnote{See \cite{Cheung:2018cwt,Hamada:2018dde,Montero:2018fns,Arkani-Hamed:2021ajd} for work related to the derivation of the mild versions, and also \cite{Bonnefoy:2019nzv} for a proposed relation between the Distance conjecture  and black hole entropy for very large black holes, which is the opposite limit to the one considered here.}, thus providing a new bottom-up rationale for them, independent of string theory. Since black hole entropy can be computed as a particular partition function (the entropy is related via a Legendre transform to the free energy, which is the logarithm of the grand canonical partition function), this is going to be the first example discussed in this paper where Swampland constraints can be derived from finiteness of quantum gravity amplitudes. We will elaborate more on this notion of finiteness of amplitudes in Section \ref{sec:finiteness}.

Our starting point is an Einstein-Maxwell-scalar system in four dimensions, with lagrangian 
\begin{equation}\label{action} S=\int d^4x \sqrt{-g}\left[R+2\vert d\phi\vert^2+\frac{1}{2g(\phi)^2} \vert F\vert^2\right].\end{equation}
The gauge coupling function $g(\phi)$ is left arbitrary, except for the fact that we demand that the gauge coupling $g(\phi)\rightarrow 0$ as $\phi\rightarrow \infty$. Hence, the infinite distance point is a also weak coupling point in which a global symmetry would be restored. In all controlled asymptotic limits in string theory, we know more: the gauge coupling behaves exponentially on the field distance as $g=\frac{1}{\sqrt{2}}e^{-a\phi}$,
with $\alpha$ some positive constant. However, the argument we will present works for general gauge coupling dependence. For the sake of clarity and for the interested reader, we work out our argument in the particular case of an exponential dependence in Appendix \ref{app:stab}.

We will be interested in electrically charged black hole solutions (no magnetic charge). If we took $g(\phi)$ constant, these would be the usual Reissner-Nordstrom solutions, that approach finite area at the horizon. When the gauge coupling depends on a scalar exponentially as above, the behavior is quite different. One finds that the gauge coupling runs to zero at the core of the black hole, which becomes very small close to extremality, forcing a parametrically large field displacement of the dilaton as we approach the horizon \cite{Garfinkle:1990qj}. For this reason, these solutions are often called ``small black holes'' in the literature \cite{Sen:1995in,Sen:1994eb}\footnote{Not to be confused with the notion of a ``small AdS black hole'', which merely refers to a black hole in Anti de Sitter space whose size is much smaller than the AdS length.}. What typically happens in stringy embeddings of these black holes in which the scalar corresponds to the dilaton field is that the effective area and the dynamics at the core are controlled by stringy effects where the EFT description breaks down. This is directly related to the large vev that the dilaton attains at the core, since it controls the string scale (i.e. the EFT cutoff) in Planck units. We will now see that:\begin{itemize}
\item The existence of these small black holes is independent of the particular dependence of $g(\phi)$, as long as it vanishes at infinite distance, and
\item They can lead to a violation of the Bekenstein bound, unless the EFT cutoff decreases exponentially on the field distance and proportionally to the gauge coupling as dictated by the Distance Conjecture and WGC.
\end{itemize}

We will now show the first point, and leave the discussion of the second point for section \ref{BHargument}. 
We are interested in a Minkowski vacuum where the asymptotic value of the dilaton is
$\phi=\phi_0$,
and consider extremal black hole solutions of electric charge $Q$. By extremal we mean that the value of the mass is the minimum one allowed before reaching a naked singularity; it is precisely these extremal solutions the ones that display the phenomenon of zero area at the core. Of course, there are also sub-extremal solutions, which have finite horizon area. Not only we can construct them, but their existence can then be inferred from physical grounds: just take a Schwarzschild black hole and add a small amount of charge. This line of thought also provides an argument that the small black hole solutions that we discuss are necessarily physical states in the theory, since one can just evaporate a sub-extremal solution until the black hole temperature is of order the cutoff.

The lagrangian \eq{action} is precisely what one would obtain in a truncation to the bosonic subsector of an $\mathcal{N}=2$ supergravity, and so these black holes can be conveniently studied by importing techniques from the attractor mechanism literature \cite{Gibbons:1996af,Denef:1998sv,DallAgata:2010ejj}.
A good ansatz for the spherically symmetric, static metric of the extremal electric solution of charge $Q$  is given by \cite{Denef:1998sv}
\beq
ds^2=-e^{2U}dt^2+e^{-2U}\left(\frac{1}{h(r)^2}\frac{d\tau^2}{\tau^4}+\frac{1}{\tau^2}d\Omega_2^2\right),
\eeq
where the coordinate $\tau$ runs from $-\infty$ at the black hole horizon to $0$ at asymptotic infinity \cite{Gibbons:1996af}. We also have an electric field turned on,
\begin{equation} F=\frac{g^2}{4\pi} Q e^{2U}\tau^2\, dr\wedge dt,\label{efield}\end{equation}
and a radial dilaton profile $\phi(r)$. As explained in \cite{Denef:1998sv}, the function $h(r)$ can be set to one without loss of generality. Then, the only independent functions in the problem are $\phi(r)$ and $U(r)$. The complete dynamics of these two is captured by the equations of motion of the following one-dimensional lagrangian \cite{Gibbons:1996af,Denef:1998sv}:
\begin{equation}\mathcal{L}_{1d}=\frac12\left(\dot{U}^2+\dot{\phi}^2\right)+g^2Q^2e^{2U},\end{equation}
(where the dots denote derivatives with respect to $\tau$) subject to a constraint (the Hamiltonian constraint),
\begin{equation}\dot{U}^2+\dot{\phi}^2-g^2Q^2e^{2U}=0.\label{eeqs0}\end{equation}
These are exactly the Newtonian equations of motion of a two-dimensional particle of zero energy moving in the potential $V_{\text{Newtonian}}=-g^2Q^2 e^{2U}$. The area of the 2-spheres is
\begin{equation} A\equiv \frac{e^{-2U}}{\tau^2}.\end{equation}
It follows that a black hole will go to zero area at the horizon if $e^{-2U}$ grows slower than $\tau^2$ as $\tau\rightarrow-\infty$. This happens if and only if the gauge coupling goes to zero as $\tau\rightarrow-\infty$. From \eq{eeqs0}, we see that
\begin{equation}\dot{U}^2\leq g^2 Q^2 e^{2U}.\label{e32455}\end{equation}
The equation of motion for $U$,
\begin{equation}\ddot{U}=2g^2 Q^2e^{2U},\label{emoU}\end{equation}
means that the acceleration is everywhere positive. Since $\dot{U}(-\infty)=0$ (because at the location of an extremal horizon we have a double zero of the scale factor), we have $\dot{U}>0$, and we can take square roots in \eqref{e32455} to yield
\begin{equation} \frac{dU}{e^{U}}\leq g Q\, d\tau,\end{equation}
and integrating,
\begin{equation} -e^{-U(0)}+e^{-U(\tau)}\leq Q \int_{\tau}^0 g(\tau)\, d\tau,\end{equation}
which rearranges, after using $U(0)=0$ due to asymptotic flatness, as
\begin{equation}A(\tau)=\frac{ e^{-2U(\tau)}}{\tau^2}\leq \frac{1}{\tau^2}\left(1+  Q \int_{\tau}^0 g(\tau)\, d\tau\right)^2.\label{56667}\end{equation}
Taking the limit $\tau\rightarrow-\infty$ gives us an upper bound on the horizon area. Taking $g(\tau)=\text{constant}$, the right hand side of \eqref{56667} asymptotes to a nonzero constant. In this case, the deep core of the black hole is described by the usual Reissner-Nordstrom solution, where indeed $A(-\infty)>0$ and the bound is saturated. In any other case (assuming $g(-\infty)\rightarrow0$), the integral in the right hand side of \eq{56667} grows more slowly than $\tau^2$, and so we must have $A(-\infty)\rightarrow 0$. 

An important ingredient in the above analysis is the assumption that the solution of the equations of motion can be fully extended all the way to $\tau\rightarrow-\infty$. That's what happens in a black hole solution; if the solution stops at a finite value of $U$, the metric will not have a double zero, and the metric will not describe a black hole. Precisely this is what happens when one tries to solve the attractor equations near a conifold point; the resulting horizonless solutions were called ``empty holes'' in \cite{Denef:2000nb}. Hence, our arguments do not necessarily hold there, which is consistent with the fact that the conifold point is at a finite distance in the moduli space and there is no tower of states becoming light. It is only at the weak coupling points associated with infinite field distance limits where we expect to be able to get trouble with remnants originating from the presence of these small black hole solutions. 

To sum up, parametrically large field variations associated with weak coupling limits are confined near small regions in space and are necessarily associated with very small black holes. This suggests that parametrically large field excursions necessarily probe the UV of the theory, as we need to go to very small regions, explaining why they are constrained by the quantum gravity completion. This was also noticed in \cite{Draper:2019utz}.

\subsection{An entropy argument for the Distance and Weak Gravity Conjectures\label{BHargument}}
Since the black holes discussed above become nearly point-like objects, we can treat them as particles and count how many different states can we construct that fit in a box of large but fixed size $L$. We allow for strong gravitational effects in localized regions inside the box, but gravity should be weakly coupled near the box boundaries. In particular, we will take the area of the box to scale as $L^2$.  

Our basic observation is that the number of distinct states that we can fit inside of the box should not grow larger than $\sim L^2$, since that would lead to a violation of entropy bounds. More concretely, a region of size $L$ cannot have more entropy than a Schwarzschild black hole of the same area\footnote{The upper bound from the Schwarzschild entropy applies to charged states as well, since they also contribute to the canonical ensemble.}. This is the same line of argument behind the derivation of the species bound in \cite{Dvali:2007hz,Dvali:2007wp}.

So how many black hole states can we fit in the box? Treating them as point-like objects, the answer is that, at sufficiently weak coupling (by tuning $\phi_0$ to a very large value), we can fit as many as we want. At very large $\phi_0$, the strength of the electric field \eqref{efield} at a given distance from the core of a singular black hole can be made arbitrarily small. Furthermore, since these black holes satisfy a BPS condition (which stems from the fact that \eq{action} is a consistent truncation of $\mathcal{N}=2$ supergravity), the leading gravitational and gauge interactions cancel each other. And since we are taking them to be pointlike, gravitational interactions depending on their geometric cross-section also become arbitrarily small. With these assumptions, any small black hole of charge $Q$ effectively counts as an additional species when computing the total entropy of the box. We wish to consider as many states as we can within the box, so we will work in the microcanonical ensemble at an energy 
\begin{equation} E\sim L\label{e322}\end{equation}
in Planck units. This is a system barely below its Schwarzschild radius; it is about to collapse into a black hole, as in \cite{Dvali:2007hz}. 

The total entropy will be a sum over the contributions of the different species, 
and it is dominated by light species with $m\ll T$ where $T$ is the temperature $\frac{1}{T}=\partial_ES$. 
Since the mass of the species of charge $Q$ is just $m=g(\phi_0) Q$, we can ensure that any number of them is light enough by going to sufficiently weak coupling\footnote{The condition that all species below a charge $Q_{\text{max}}$ are light is that $Q_{\text{max}} g_0 \leq T\sim\frac{1}{(Q_{\text{max}} A^{1/4})}$, where $A\sim L^2$ is the area of the box. As described in the main text, we will impose $Q_{\text{max}}<A$ due to entropy bounds, which leads to $g_0<A^{-3/2}$. The equivalent bound $A\sim g_0^{2/3}$ is precisely the size of the quantum gravity cutoff associated with a tower of particles \cite{Heidenreich:2017sim,Grimm:2018ohb}; therefore, to make the argument in this Section, we must consider the smallest box that could possibly make sense before local physics breaks down.}. For the sake of concreteness, let us consider the case of four dimensional space-time (although the argument is valid for any dimension), so that we can 
 use the standard relations 
 \begin{equation} S=N_{\text{species}} T^3\, L^3,\quad E= N_{\text{species}}T^4\, L^3.\end{equation}
Together with the condition \eq{e322}, one obtains that $S=N_{\text{species}}^{1/4}E^{3/4}L^{3/4}=N_{\text{species}}^{1/4} L^{3/2}$, and so, imposing the entropy bound that this is below the area, $S\leq L^2$ leads to
\begin{equation} N_{\text{species}}=Q_{\text{max}}\lesssim L^2, \label{entropy}\end {equation}
where $Q_{\text{max}}$ is the charge of the largest light black hole species under consideration, and we have taken the conservative assumption that $N_{\text{species}}=1$ for each value of the charge. The bound \eq{entropy} will surely be violated in a theory in which the small black holes are exactly pointlike, since we can take $Q_{\text{max}}$ arbitrarily large.

It is instructive to consider why \eq{entropy} is not violated in an Einstein-Maxwell theory, without a dilaton coupling. There, the extremal solutions are Reissner-Nordstrom black holes of finite size; as a result, one just cannot fit an arbitrarily large number of species in a box of size $L$. Rather, $Q_{\text{max}}\propto L$, since the charge of an extremal RN black hole is proportional to its radius, and \eq{entropy} is satisfied.

Going back to the Einstein-Maxwell-dilaton theory, we will now see how the contradiction with \eq{entropy} is avoided by taking into account the fact that small black holes have a nonzero effective size, given by the cutoff $\Lambda$ of the EFT. More concretely, keeping the gradient of the scalar field below the EFT cutoff (so that the black hole solution makes sense) forces us to stop at a certain distance from the horizon, such that the black holes have some effective finite area which may increase with their charge $Q$.
On the other hand, if the cutoff changes sufficiently quickly with the field displacement, the black holes will grow quickly in size, and we will not be able to fit too many of them in a box of size $L^2$, thereby avoiding the contradiction.

We will use the gradient of the scalar fields as a proxy for when the EFT description should break down. First, the gradient of the scalar field can achieve arbitrarily high values in the small black hole solution. This is because the equation of motion for $\phi$,
\begin{equation} \ddot{\phi}=\frac{dg^2}{d\phi} Q^2\,e^{2U},\end{equation}
can be integrated to give, using that $\dot{\phi}(0)=0$ and that $e^{2U}$ is a monotonic function (which follows from the equation of motion \eq{emoU}), that 
\begin{equation} \dot{\phi}^2\geq \Delta g^2 Q^2 e^{2U},\end{equation}
where $\Delta g^2$ is the change in $g^2$ from its asymptotic value. It is a finite, bounded quantity.
The gradient of $\phi$ is then 
\begin{equation} \vert d\phi\vert^2= \tau^4e^{2U} \dot{\phi}^2 \geq \frac{\Delta g^2 Q^2}{A^2},\end{equation}
which indeed diverges close to the core as the area goes to zero. The EFT breaks down whenever the scalar field gradients are of order the cutoff, so that
\begin{equation} \tau^4e^{2U} \dot{\phi}^2\sim \Lambda^2.\label{eww3}\end{equation}
Rearranging, and using again the Hamiltonian constraint \eq{eeqs0}, one finds that
\begin{equation} \frac{\Lambda^2e^{-2U}}{\tau^4}=\frac{A}{\tau^2}\Lambda^2\leq \dot{\phi}^2 \leq g^2 Q^2\, e^{2U},\end{equation}
which in turn can be rearranged to give
\begin{equation} \Lambda\leq g \frac{Q}{A}.\end{equation}
Now, \eq{entropy} tells us that the maximum $Q$ that we can consider in a box of size $L$, and area $A\sim L^2$, is lower than $A$. So $Q/A<1$, and we obtain 
\begin{equation} \Lambda\leq g,\end{equation}
in Planck units, which is precisely the magnetic version of the WGC \cite{Arkani-Hamed:2006emk}.

To our knowledge, all black hole arguments behind the WGC are based on black hole stability: the existence of a WGC particle is required to allow extremal black holes to decay. As a result, one can only argue for a mild version of the WGC \cite{ArkaniHamed:2005yv} i.e. that there has to be one superextremal light state. 

By contrast, our argument for the WGC is based on the finiteness of entropy, or absence of charged remnants, and requires the existence of infinitely many charged states becoming light, since the local EFT description must break down (integrating a finite number of charged states will not affect our considerations). Hence, we have actually argued for a Tower version of the WGC, which is stronger than the mild version typically derived from the typical black hole stability argument. The argument given here can be regarded as a more quantitative version of the original motivation for the WGC given in \cite{Arkani-Hamed:2006emk}; by considering small black holes, the trouble with small gauge couplings outlined there can be made precise.

Whenever the gauge coupling behaves exponentially on the proper field distance, the above cutoff will also decrease exponentially on the distance, as dictated by the Distance Conjecture,
\beq
\Lambda\leq g \sim e^{-\alpha \phi}\ .
\eeq
This is the behavior for the gauge couplings found in all the infinite distance limits of string compactifications known so far. Hence, assuming this exponential behavior of the gauge coupling, our entropy argument also reproduces the Distance Conjecture since, as explained above, the cutoff signals the presence of infinitely many states becoming light. Nevertheless, it is fair to ask to what extent this exponential behavior is just a lamppost effect, since the above black hole argument is not sensitive to it; it always holds as long as the gauge coupling vanishes asymptotically. A bottom-up argument in favor of this universal exponential behavior for the gauge coupling comes from the Emergence proposal \cite{Harlow:2015lma,Heidenreich:2017sim,Grimm:2018ohb,Heidenreich:2018kpg,Corvilain:2018lgw,Palti:2019pca}. It was argued in \cite{Heidenreich:2017sim,Grimm:2018ohb} that such behavior emerges universally from quantum corrections of integrating out the tower of states, in a similar way that the emergence of the field metric that we will discuss in Section \ref{sec:emergence}.

Let us also remark that an important assumption in our argument was $N_{\text{species}}\geq1$ for each $Q$, i.e. we assumed there is at least one black hole species for each value of the charge contributing to \eq{entropy}.
One may worry that for some values of $Q$ the black holes decay by emission of charged particles, and the problem above is avoided. This worry is averted by the fact that we do not need the small black holes to be exactly stable; it is enough if they are sufficiently long-lived, with a lifetime significantly larger than the box size $L$. That this is indeed the case is justified in appendix \ref{app:stab}.

We will now comment briefly on the stringy embedding of our results. These small black holes have a natural lift to electrically charged BPS black holes in 4d $\mathcal{N}=2$ theories arising from compactifying Type IIB on a Calabi-Yau threefold, for instance. The field displacement can then be mapped to a field distance trajectory approaching an infinite distance limit in the complex structure moduli space. In such a case, the small black holes  correspond to the usual modes of the infinite tower of BPS states becoming light, arising from wrapping D3-branes on electric 3-cycles. They are both extremal and feel no force (i.e. they saturate the WGC and the Repulsive Force Condition \cite{Palti:2017elp,Heidenreich:2019zkl}), implying that indeed the exponential rate of the tower is determined by the behavior of the gauge coupling, as explicitly shown in \cite{Lee:2018spm,Gendler:2020dfp} using the asymptotic geometry of the moduli space. In other words, the bottom-up result for the exponential rate of the cutoff, and consequently, of the mass of the tower, can be matched to the general properties of Calabi-Yau moduli spaces. 
This is just an example of a more general phenomenon; whenever there is a gauge coupling vanishing at infinite distance, there is a tower of states satisfying both the Distance Conjecture and the WGC such that the exponential mass rate is bounded by the black hole extremality bound \cite{Lee:2018spm,Gendler:2020dfp}. Our argument provides a bottom-up rationale for this empirical result in string theory.

Finally, one could also wonder if we can have infinite distance limits which do not correspond to weak coupling points, such that the above black hole argument does not apply. Currently, all string theory examples have a vanishing p-form gauge coupling at infinite field distance. This was proposed to be a general feature in \cite{Gendler:2020dfp}. This expectation matches with the fact that the Distance Conjecture is strongly linked to the manifestation of dualities at asymptotic limits \cite{Ooguri:2006in}, so that the tower hints a new weakly coupled description of the theory. An interesting extension of our work would be to generalize our entropy arguments to black $p$-branes, so that they can be used for more general infinite distance limits associated with weak coupling points for $p$-form gauge fields.

\section{UV Compactness and Emergence proposal} \label{sec:emergence}
In the previous Section, we have seen how an argument based on entropy bounds can be given for the Distance Conjecture, close to any infinite distance limit where a gauge coupling goes to zero. We will now give a different class of arguments which, even though slightly weaker, have the benefit of working in a general asymptotic limit, even if a gauge coupling does not vanish there. They are also based on the finiteness of entropy.
\subsection{UV Compactness\label{sec:compact}}

The Bekenstein-Hawking entropy of black holes is obtained from the gravitational partition function in the semiclassical approximation \cite{Gibbons:1976ue}. In this sense, the finiteness of the Bekenstein-Hawking entropy implies the finiteness of the partition function.
We will now argue that a breakdown of low energy EFTs for large field excursions can be derived by assuming the finiteness of the partition function.
This provides a bottom-up argument for a weaker version of the Distance Conjecture, namely that EFTs can only be valid for a finite variation of the scalar fields.

We will illustrate our arguments with a simple toy model: the $d$-dimensional free-scalar theory,  with an EFT cutoff $\Lambda$.
The Lagrangian is 
\begin{align}
\mathcal{L} = \frac{1}{2}\left(\partial_\mu\varphi\right)^2,
\label{Eq:d-dim_Lagrangian}\end{align}
where $\varphi$ is a non-compact scalar.
We can quantize the theory on some spatial manifold $X_{d-1}$, and consider its partition function
\begin{align}
Z=\mathrm{Tr} \,e^{-\beta H},
\label{Eq:partition_function}\end{align}
where $\beta$ is the inverse temperature, and $H$ is the Hamiltonian corresponding to \eqref{Eq:d-dim_Lagrangian}. 

In the previous Section, we have assumed a sharp bound on the entropy.  We will now take the much milder point of view that the partition function \eq{Eq:partition_function} does not have to satisfy a particular bound, but must at least be finite in a consistent quantum theory of gravity.  
In Section~\ref{sec:finiteness}, we will give more general arguments for the assumption.

In particular, we require $Z$ is finite when the spatial manifold is taken to be compact, for instance, $X_{d-1}=T^{d-1}$ or $S^{d-1}$.
What is special about compactification to one and two dimensions is that the asymptotic values of moduli are not fixed; the zero modes of the scalar field are dynamical, and the partition function includes an integral over them.

After the compactification, the one dimensional Lagrangian is
\begin{align}
&L = \frac{V_{d-1}}{2}\dot{\varphi}^2= \frac{1}{2}\dot{\phi}^2,
\end{align}
where $V_{d-1}$ is the volume of $X_{d-1}$, and $\phi=\varphi\sqrt{V_{d-1}}$.
The spectrum of one dimensional theory is determined by studying the quantum mechanics Hamiltonian
\begin{align}
H= \frac{1}{2}\hat{p}^2,
\label{Eq:Hamilatonian}\end{align}
where $\hat{p}$ is the momentum operator conjugate to $\phi$.

We can see that the partition function~\eqref{Eq:partition_function} diverges. The Hamiltonian~\eqref{Eq:Hamilatonian} is diagonalized by the eigenstates of $\hat{p}$ as $\hat{p}|p\rangle=p|p\rangle, H|p\rangle = E |p\rangle$ where $E=p^2/2$.
Explicitly, we can provide a lower bound on the partition function as
\begin{align}
Z > \sum_n^{E_n=\Lambda} e^{-\beta E_n} 
> \sum_n^{E_n=\Lambda} e^{-\beta \Lambda} 
=N(\Lambda) \,e^{-\beta \Lambda} ,
\label{Eq:divergence}\end{align}
where $N(\Lambda)$ is the number of states whose energy is less than $\Lambda$. Here we have taken the conservative approach that we do not include the contributions of those states above the cutoff $\Lambda$, so we obtain a lower bound. On the other hand, $N(\Lambda)$ is actually infinite, due to the continuum of eigenstates above the ground state. 
Therefore, the partition function diverges, and the finiteness requirement is violated.

We will now illustrate a way to avoid the divergent partition function. It is to make $\phi$ be a compact scalar. 
Let us denote the periodicity of $\phi$ by $\phi_\text{max}$, $\phi\simeq \phi+\phi_\text{max}$.
Then, the eigenvalues of $\hat{p}$ are discrete, and the partition function is computed as
\begin{align}
Z=\sum_{n=-\phi_\text{max}\sqrt{\frac{\Lambda}{2\pi^2}}}^{\phi_\text{max}\sqrt{\frac{\Lambda}{2\pi^2}}} \,e^{-2\beta n^2\pi^2/\phi_\text{max}^2}.
\label{Eq:periodic}\end{align}
This is clearly finite since it is the sum of a finite number of terms.

By taking $\phi_\text{max}$ to be large, an effective non-compact scalar is obtained.
It is instructive to see what happens when we take this limit.
The partition function \eqref{Eq:periodic} becomes
\begin{align}
Z\simeq \frac{\phi_\text{max}}{\sqrt{2}\,\pi} \int^{\sqrt{\Lambda}}_{-\sqrt{\Lambda}} dp\, e^{-\beta p^2}
=\frac{\phi_\text{max}}{\sqrt{2\pi\beta}}\,\mathrm{erf}\left(\sqrt{\beta\Lambda}\right),
\end{align}
where $\mathrm{erf}$ is the error function.
It diverges for $\phi_\text{max}\to\infty$, in accordance with the argument around \eqref{Eq:divergence}.
The only way to keep a finite partition function is that $\Lambda$ is a function of $\phi_\text{max}$ and satisfies
\begin{align}
\Lambda\left(\phi_\text{max}\right) \leq \frac{1}{\phi_\text{max}^2}\quad \text{for} \quad  \phi_\text{max}\to\infty.
\label{Eq:bound}\end{align}
Namely, the EFT cutoff scale $\Lambda$ should decay equal or faster than $\phi_\text{max}^{-2}$.
This can be viewed as a weaker version of the Distance Conjecture.
For fixed cutoff $\Lambda$, the EFTs are valid only a finite diameter of the scalar fields.
The power-law decay we have obtained here is weaker than the exponential decay demanded by the Distance Conjecture.

\subsection{Emergence proposal}

The above argument from the finiteness of amplitudes only implies a weaker version of the Distance Conjecture, i.e. it accounts for the EFT breakdown due to fields becoming light, but it does not reproduce the exponential behavior of their masses. A tantalizing possibility is that this exponential behavior is a consequence of the infinite tower of states becoming light upon taking into account quantum corrections induced by the tower on the field metric. This is the rationale behind the Emergence proposal \cite{Grimm:2018ohb,Heidenreich:2018kpg}, in which the infinite distance (and the exponential behavior of the cutoff) emerge upon integrating out the infinite tower of states.

The basic logic of the Emergence proposal \cite{Palti:2019pca,vanBeest:2021lhn} is that dynamical properties of a full quantum gravity theory are encoded in the kinematic properties of a free particle spectrum.  In particular, emergence dictates that, in a weakly coupled theory, all kinetic terms (including moduli fields) emerge in the IR after integrating out states. In some vague sense, yet to be made precise, emergence suggests that there is some notion of configuration/moduli space of a quantum theory of gravity in the UV, which is compact.   Hence, Emergence flips this logic of the Distance Conjecture, and shows how the infinite distance limits in moduli space are ``emergent'', an artifact of the IR description, and can only appear when there is an infinite tower of states becoming massless at some point of the moduli space.

Interestingly, the finiteness of the partition function in the previous Subsection can be used to argue for the main assumption behind the Emergence proposal, i.e. that the moduli space is compact in the UV and the infinite field distance only emerges upon taking quantum corrections in the IR from integrating out states\footnote{See \cite{Stout:2021ubb} for a recent approach that explores instead the relationship of emergence at infinite distance to information theory}.   As we explained above, a non-compact moduli space would imply a divergent partition function upon compactifying to 2 or lower dimensions, which is problematic in quantum gravity. Once we establish UV compactness, the infinite distance in the IR field space and the exponential behavior of the cutoff can be reproduced from loop corrections of integrating out a tower of states becoming massless, as we briefly review in the following.   We will closely follow the computation shown in \cite{Corvilain:2018lgw} (see also \cite{Grimm:2018ohb,Heidenreich:2018kpg}).

Let us consider a $d$-dimensional theory containing a tower of states $\psi_n$ whose mass is parametrized by a modulus $\phi$, i.e. there would be terms
\begin{equation}\mathcal{L}\supset \sum_n\left(\frac12 \vert \partial\psi_n\vert^2- \frac12 m^2_n(\phi) (\psi_n)^2\right)\label{e32},\end{equation}
if the theory was Lagrangian.
At a particular value of the modulus $\phi=\phi_0$, one can expand 
\begin{equation}\phi=\phi_0+\delta\phi,\end{equation}
and then \eqref{e32} gives trilinear couplings between the scalar $\psi_n$ and $\phi$ of the form
\begin{equation} \lambda \psi_n^2\delta\phi,\quad \lambda\equiv  m \partial_\phi m\vert_{\phi=\phi_0}.\end{equation} The one-loop contribution to the $\phi$ propagator
can be readily computed using Feynman rules, as (external momentum is $p$, internal momentum is $q$)
\begin{equation} \mathcal{A}_n(p^2)=\lambda^2\int \frac{d^dq}{(2\pi)^d} \frac{1}{q^2+m_n^2} \frac{1}{(p-q)^2+m_n^2}.\end{equation}
so that the correction to the field metric is given by
\begin{equation} \delta g_{\phi\phi}=\sum_n \left.\frac{\partial\mathcal{A}_n}{\partial p^2}\right\vert_{p^2=0}=\sum_n\lambda_n^2\frac{\Omega_{d-1}}{(2\pi)^d} \int \frac{ q^{d-1} \left(3 q^2-m_n^2\right)}{\left(m_n^2+q^2\right)^4}.\end{equation}
where we have included a sum over all the particles.
The contribution from each particle diverges in $d\geq 5$, so we will cut off the momentum integrals at a scale $\Lambda$. Doing so, we get
\begin{equation} \delta g_{\phi\phi}= \sum_n \lambda_n^2\left(\frac{1}{48} \pi  (d-4) (d-2) (2 d-3) m_n^{d-6} \csc \left(\frac{\pi d}{2}\right)+ \sum_0^{k\leq d} c_k\left(\frac{m_n^{2k}}{\Lambda^{6-d+2k}}\right)\right),\label{we2}\end{equation}
where the $c_k$ are finite coefficients one can compute explicitly and that only depend on the space-time dimension. 
The last terms in \eqref{we2} vanish in the $\Lambda\rightarrow\infty$ limit for $d<6$, but for $d\geq6$ they are relevant and  $c_1$ gives the leading divergent contribution. In general, in $d$ dimensions, we have
\begin{equation} \label{corr}\delta g_{\phi\phi}\sim \sum_n \lambda_n^2  \left\{\begin{array}{lr}
       m_n^{d-6}, & \text{for } d\leq 6\\
        \Lambda^{d-6}& \text{for } d>6
        \end{array}\right.
        \end{equation}
In what follows, we focus on $d\leq 6$, although the results also apply for $d>6$ since the integral will be dominated by the UV modes with $m\sim \Lambda$ as it will become clear later. 
This is divergent due to the sum over the infinite tower, but we should stop counting states at the quantum gravity scale, where quantum gravitational effects become important and no local effective field theory description makes sense anymore. In Planck units, this is given by the species bound \cite{ArkaniHamed:2005yv,Dvali:2007wp,Dvali:2007hz,Dvali:2010vm}
\begin{equation}N\sim \frac{1}{\Lambda^{d-2}},\label{ew2}\end{equation}
where $N$ is effective the number of weakly interacting field species until the quantum gravity scale $\Lambda$. In the case of a tower of particles, $N$ is given by  
the number of states in the tower such that their mass $m_n$ is lower than $\Lambda$.
 We will now assume that the masses $m_n$ depend homogeneously on some mass scale $\Delta m$, i.e.
\begin{equation}
 m_n=f(n) \Delta m,
 \end{equation}
where $f(n)$ is an arbitrary function. The quantum correction to the  field space metric for $d\leq 6$ then reads
\begin{equation} \delta g_{\phi\phi}\simeq  \sum^N_{n=1} m_n^{d-4}\, (\partial_\phi m_n)^2= (\Delta m)^{d-4}( \partial_\phi \Delta m)^2 \left[\sum^N_{n=1}f(n)^{d-2}\right],\label{ewe3}\end{equation}
where $N$ itself depends on $\Lambda$ via \eqref{ew2}.

 Let us first consider the case of an infinite tower of states of equidistant mass, so that $f(n)=n$ (e.g. a KK tower).
The number $N$ of states below the species scale $\Lambda$ is given by
\beq
N=\frac{\Lambda}{\Delta m}
\eeq
which combined with \eqref{ew2} gives
\beq
\Lambda=(\Delta m)^{1/(d-1)}\ , \quad N=( \Delta m)^{-\frac{d-2}{d-1}} 
\eeq
Plugging this back into \eqref{ewe3}, we obtain
\begin{equation} 
\delta g_{\phi\phi}\simeq (\Delta m)^{d-4} (\partial_\phi \Delta m)^2 N^{d-1}=\left(\frac{\partial_\phi \Delta m}{\Delta m}\right)^2
\end{equation}
which leads to a logarithmic dependence on the distance 
\begin{equation} 
s\sim\int d(\log\Delta m)=\log \Delta m
\end{equation}
as predicted by the Distance Conjecture.  Notice that this is a Swampland statement because $M_{pl}$ has entered the calculation though the species bound (we have set $M_{pl}=1$).  Thus, we see than when a tower of states becomes light $\Delta m\rightarrow0$ at some point of the moduli space, quantum effects reproduce the infinite distance. Remarkably, regardless of the specific dependence of $\Delta m$ with $\phi$, the mass of the tower will behave exponentially in terms of the quantum corrected distance.  

The same parametric dependence appears if $f(n)$ is some arbitrary monomial on $n$ (e.g $f(n)=\sqrt{n}$ for a tower of string excitation modes) or even if the states are roughly degenerate in mass so $f(n)=const$. In the latter case, the mass of the states also sets the quantum gravity cutoff $m_n=\Delta m\simeq \Lambda$, obtaining
\begin{equation} 
\delta g_{\phi\phi}\simeq  N m^{d-2} (\partial_\phi m)^2 =\left(\frac{\partial_\phi  m}{ m}\right)^2
\end{equation}
which leads again to the logarithmic dependence on the cutoff
\begin{equation} 
s\sim\log m\sim \log \Lambda \ .
\end{equation}
One can also check that the same result is obtained in $d\geq 6$ using \eqref{corr}.

More generally, let us not assume anything on $m_n$, other than the assumption that the integral in \eqref{ewe3} can be approximated as 

\begin{equation} \sum^N_{n=1} m_n^{d-4}\, (\partial_\phi m_n)^2 \approx N \cdot \left(m_n^{d-4}\, (\partial_\phi m_n)^2\right)_{\text{max}},\label{wq2}\end{equation}
where $(\ldots)_{\text{max}}$ is the maximum value of the integrand. This is true of any monotonic function that grows sufficiently fast; for instance, it is true for $f(n)=n^\alpha$ above for any $\alpha$. In that case, the species bound sets $m_N\sim \Lambda$, and so \eqref{wq2} becomes
\begin{equation} \sum^N_{n=1} m_n^{d-4}\, (\partial_\phi m_n)^2 \approx \left(\frac{\partial \Lambda}{\Lambda}\right)^2.\label{wq3}\end{equation}
which again produces a cutoff that decreases exponentially with proper field distance, thus, providing a derivation of the Distance Conjecture.

We end this Subsection with what we find is an interesting consistency check of the Emergence proposal. Let us go back to the case of an equidistant tower. 
The above considerations, loop diagrams, etc work in any number of dimensions, in particular, also in 2d. The number of states included in the sums above is
\begin{equation} N=(\Delta m)^{-\frac{d-2}{d-1}},\label{brylig}\end{equation}
and for $d>2$ it diverges as $\Delta m\rightarrow0$, so to get infinite distance we need an infinite number of states. However, for $d=2$, \eqref{brylig} $N$ turns out to be independent of $\Delta m$. This suggests that, in $d=2$ (and also in $d=1$), it is possible to achieve infinite distance with only a finite number of states becoming light. The above loop computation yields the logarithmic divergence of the distance even if $N=1$ for $d=2$. 

In fact, we know examples of this kind of emergence of infinite distance in 2d QFT from worldsheet perturbation theory, for example as seen in \cite{Witten:1994tz,Witten:1995zh}. The worldsheet sigma model of a small heterotic instanton involves two scalars $X$ and $\phi$, with a potential
\begin{equation} V=\frac18 X^2\, \phi^2\label{e33}\end{equation}
and a metric such that the point $X=\phi=0$ where the two branches meet is at a finite distance. Suppose we give a vev to $X$, and consider the theory as $\langle X \rangle\rightarrow0$. As we do this, equation \eqref{e33} tells us that the massive field $\phi$ is becoming massless. The loop computation above suggests hence that the region $X=0$ is actually at the end of a long tube, and ends up being at infinite distance. This is exactly what was found in \cite{Witten:1994tz}: The $X=\phi=0$ is pushed to an infinite distance, with the two branches disconnecting.

To summarize, emergence makes the post-diction that in low dimensions it should be possible to obtain infinite distance limits with only a finite number of particles becoming massless, which is verified in known examples coming from string perturbation theory. This also explains why in the argument of Subsection \ref{sec:compact} we could not conclude that there had to be an infinite number of particles becoming light to generate the infinite distance; as illustrated above, a single particle becoming light is enough, in low spacetime dimension.

\subsection{Instantons and emergence}  

The microscopic nature of the tower of states is not specified by the Distance Conjecture. They can correspond to KK particles but also to excitation modes of extended objects, as long as they are weakly coupled. In all string theory examples studied so far, the tower consists either of KK-like particles or string excitation modes, consistent with the Emergent String Conjecture \cite{Lee:2019wij}. An interesting question is whether we could have instead a tower of instantons, as has been studied in \cite{Marchesano:2019ifh,Baume:2019sry}. Here, we want to comment on this possibility and the interplay with the above Emergence proposal.

In certain cases, it can happen that a tower of instantons becomes ``light'' (i.e. their action goes to zero) at certain (limit) points of the moduli space. These instantons can then correct the classical field metric, and even make the distance finite even though it originally seemed infinite. For instance, \cite{Marchesano:2019ifh} studied certain trajectories in the Kahler moduli space of Type IIB compactified on Calabi-Yau threefolds along which instantons become important and correct the classical field metric. The point at which the instanton action vanishes was at an infinite distance at the classical level, but becomes finite distance upon taking into account the instanton corrections to the metric.   An even simpler example is given by the type IIA compactifications on Calabi-Yau threefolds and considering the small volume limit.
Classically this is at an infinite distance, but from mirror symmetry we know that this point is dual to a conifold point of the mirror \cite{Candelas:1989ug,Candelas:1990rm}, which is at a finite distance.  Here the infinite tower of ``light'' worldsheet instantons modifies the classical metric dramatically to make the distance finite.

 \begin{figure}[ht!]
	\centering{}\includegraphics[scale=0.40]{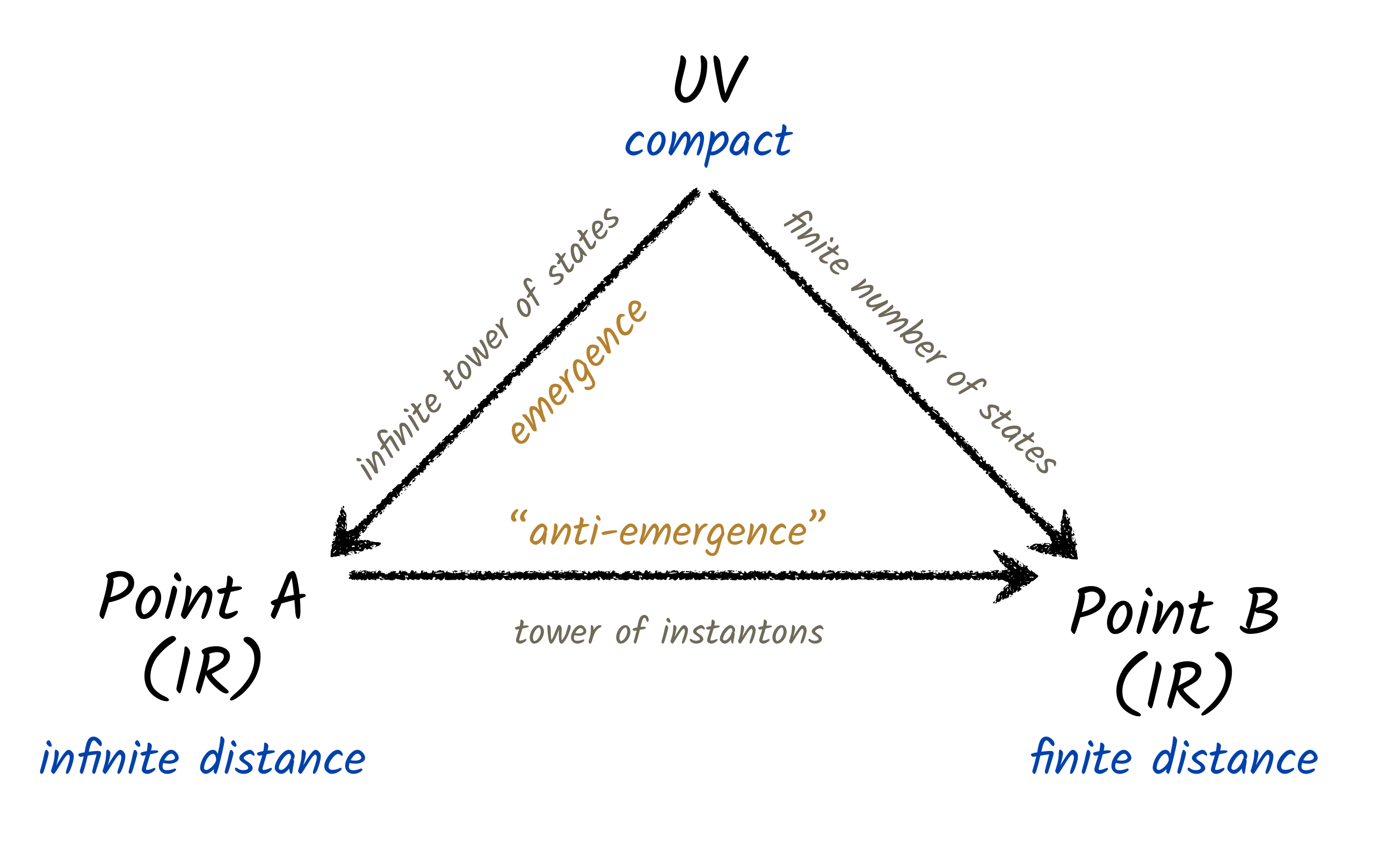} \
	\caption{Emergence from tower of particles or instantons.}
	\label{fig:emerg}
\end{figure}
Naively, these cases seem to be examples of some sort of ``\emph{anti}-emergence'', since an infinite distance becomes finite after taking into account corrections from the tower (of instantons). However, we believe the proper interpretation of these examples is as follows. Let us start with a compact UV, as required in the previous Subsection. If a tower of states becomes massless at a certain point of the moduli space (point A in figure \ref{fig:emerg}), quantum corrections from integrating out the tower will generate the infinite distance from the IR EFT perspective. We can now move to a different point in the moduli space (point B) in which there is no tower of states becoming massless. Since the moduli space was compact in the UV, it is clear that this point should be at a finite distance. 
However, we may also extrapolate the expression of the metric computed at A to a neighborhood of point B; such an extrapolation is often possible due to supersymmetry. Unlike the fundamental UV or IR metrics discussed above, the extrapolated metric is not fundamental, and may very well put B at an infinite distance. To compute the actual IR metric we may have to include corrections coming from towers of instantons or particles, which allows us to recover the correct IR metric.

Hence, the \emph{anti}-emergence induced by instantons is not a consequence of an RG flow, but an artifact arising from extrapolating the IR result at one point of moduli space to another beyond its regime of validity, as represented in figure \ref{fig:emerg}. Furthermore, the infinite tower of instantons only makes sense at point A, while at point B (where they induce the finite distance) there is typically only a finite number of bound states (i.e. there is only a finite number of non-vanishing GV invariants \cite{Gopakumar:1998ii,Gopakumar:1998ki,Gopakumar:1998jq}).

As a concrete example, one can take the IIA/IIB case discussed above, where point A is the large volume IIA description of Calabi-Yau structure and provides the classical contribution at point B which is best described by IIB on the mirror (leading to conifold point). The statement of anti-emergence is just the observation that the classical expansion around point A is very bad at point B, and receives significant quantum corrections. Stated like this, ``instanton emergence'' is clearly not a fundamental phenomenon; indeed, in the mirror perspective, already the classical metric gives the correct behavior, without ever having to invoke quantum effects.

Finally, we remark that instantons are formally charged under exotic $(-1)$-form global symmetries, a relatively novel notion in the literature \cite{Cordova:2019jnf,McNamara:2020uza,Tanizaki:2019rbk,Heidenreich:2020pkc}. $(-1)$-form symmetries share many common features with ordinary symmetries, but are qualitatively different. We have just argued that instantons are not a duality-invariant notion, and whether they are present or not depends on a particular choice of semiclassical expansion parameter. This suggests that $(-1)$-form symmetries might similarly be a frame-dependent notion. For instance, the associated topological operators can only be properly defined by extending the theory to the space of field configurations \cite{Heidenreich:2020pkc}, a notion that we only know how to make sense of semiclassically.

\subsection{Asymptotic scalar potential}

We have analyzed so far the Distance Conjecture for exactly flat moduli spaces. However, the conjecture should also hold in the presence of a scalar potential as long as the potential energy remains below the cutoff along the infinite distance path. This is necessary by the consistency of the conjecture at any energy scale (see e.g. the discussion in \cite{Calderon-Infante:2020dhm}) and has been checked in several string theory setups \cite{Baume:2016psm,Valenzuela:2016yny,Grimm:2019ixq}. We can also justify this based on the finiteness arguments given in this paper, as we explain in the following.

Consider the setup of Section \ref{sec:compact} of a d-dimensional scalar theory compactified to one dimensions. We can add a d-dimensional scalar potential such that the 1D Lagrangian becomes
\beq
L=\frac12 \dot{\phi}^2-V(\phi)
\eeq
where the potential goes to zero $V(\phi)\rightarrow 0$ as $\phi\rightarrow \infty$. The Hamiltonian still has infinitely many eigenstates in a finite window around zero energy, since one still has a continuum of scattering states. Hence, the EFT description must fail at some cutoff decreasing with the field range at least as \eqref{Eq:bound}. The situation would be different if the potential diverged asymptotically, so that the spectrum becomes discrete. However, in that case, the EFT does not have any infinite distance trajectory since this is obstructed by the potential, so there is no problem to start with.

We could wonder whether the mere existence of an infinite tower of states becoming light is only consistent with certain behaviors of the potential. In fact, one can argue that the potential cannot asymptote to a positive value at infinite field distance, since this would be in contradiction with the Higuchi bound \cite{Higuchi:1986py} (with the mild assumption that the tower contains some higher spin states which is the case in all known string theory examples, either because of KK copies of the graviton or because there is a tower of string excitation modes). If the vacuum energy remains positive when approaching the infinite distance limit, the Higuchi bound implies that
\beq
\label{higuchi}
H< m_{\rm tower}
\eeq
where $H$ is the Hubble scale. Assuming the infinite tower becomes massless at an infinite distance (which is not obvious, due to effects of the dS curvature \cite{Noumi:2019ohm,Kato:2021rdz}), the Hubble scale must, therefore, go to zero (or to negative values) to satisfy \eqref{higuchi}. Furthermore, if the cosmological constant scales in Planck units with the mass of the tower as
\beq
\label{LM}
\Lambda_{cc} =H^2 \sim m_{\rm tower}^\alpha
\eeq
then the Higuchi bound implies $\alpha>2$. This is the argument given in \cite{Lust:2019zwm} in favor of a bound for the exponent of the AdS Distance Conjecture; see \cite{Scalisi:2019gfv} for a similar approach, using the Distance Conjecture and positive vacuum energy, to put an upper bound in the inflaton field range. 

A tantalizing possibility is that the potential itself also emerges from integrating out the tower of states as suggested by the Emergence proposal. In fact, this is quite natural in the context of flux-induced scalar potentials where the potential can be re-written as the gauge kinetic function of $(d-1)$-form gauge fields in $d$ dimensions, see e.g. \cite{Marchesano:2014mla,Bielleman:2015ina}. This gauge kinetic function may then emerge from integrating out the towers of states, 
in the same way that the field metric or the 1-form gauge couplings can be interpreted as emerging from quantum corrections from the tower \cite{Heidenreich:2017sim,Grimm:2018ohb} (and reproducing this way the Distance Conjecture and WGC), see also comments in \cite{Font:2019cxq}. 

The one-loop Casimir potential from integrating out particles scales as $m^d$, so $\alpha=d$ is a natural power to expect in those cases, as remarked in \cite{Rudelius:2021oaz,Gonzalo:2021fma}.  

The relative scaling between the leading potential and the mass of the tower of states has been studied e.g. in asymptotic limits of $d=4$ $N=1$ string compactifications in \cite{Lanza:2020qmt,Lanza:2021udy}, where all examples satisfied $2\leq\alpha\leq 6$.  
 These bounds have an interpretation in terms of the tension $T$ of the membrane charged under the 3-form gauge field generating the potential; they guarantee that the membrane can be described at the semiclassical level, i.e. $T^{1/3}\geq m_{\rm tower}\geq T/M_p^2$ where $T^2\sim V$ in Planck units. Intriguingly, for the cases with $\alpha> 4$, the above estimation of the Casimir potential would suggest that quantum corrections should dominate over the classical flux-induced piece. Interestingly, this can also be used to translate bounds on the exponential rate $c$ of the potential to bounds on the exponent, since $2\lambda_{\rm tower}\leq c\leq \alpha_{max}\lambda_{\rm tower}$. For instance, using the bound for the tower found in 4D Calabi-Yau compactifications, which is $\lambda_{\rm tower}\geq 1/\sqrt{6}$ \cite{Gendler:2020dfp}, the above correlation implies that $c\geq \sqrt{2/3}$, recovering the TCC bound \cite{Bedroya:2019snp} as pointed out in \cite{Andriot:2020lea}. However, this minimum value for $c$ has not been explicitly found yet in string theory models as remarked in \cite{Rudelius:2021azq}, where a stronger bound was proposed implying always decelerated cosmologies in the asymptotic limits.

Finally, we would like to comment on the intriguing possibility that the potential is generated only non-perturbatively. For instance, if it is generated by nonperturbative effects, we have $V\sim e^{c/g^2}\sim e^{-cM_p^2/\Lambda^2}$, where $\Lambda$ is the magnetic WGC cutoff. This potentially is a very exciting possibility for our universe, since if we use the observed value of $g^2$ for the electromagnetic gauge coupling, and take $c$ of order one, one can attain a vacuum energy of the same order of magnitude as the experimental one. However, this form of the potential  is incompatible with \eqref{LM} since the mass of the tower of states  would have to be of order $\sim {\rm exp}(-a/\Lambda^2)$ instead of $\Lambda$. This suggests that this scenario can only be engineered away from the strict infinite distance boundaries of the moduli space, unless  all perturbative corrections from the tower to the potential somehow magically cancel, so that $V$ does not scale like a power of the mass in the tower.

\section{Finiteness of vacua} \label{sec:vacua}

There is a common lore that the string landscape of consistent vacua should be finite (see e.g. \cite{Acharya:2006zw}), as promoted to a Swampland conjecture in \cite{Vafa:2005ui}. In this Section, we explain how this connects with the Distance Conjecture and the more general notion of the finiteness of partition functions or more general quantum gravity amplitudes.

First of all, we should define in a more precise way what we mean with \emph{vacuum} in this context. On the one hand, theories with  8 or more supercharges have exact moduli spaces, where there is a vacuum for each value of the field; as a result, the number of vacua is clearly infinite, but in a boring way. We will ``mod out'' by moduli spaces, and consider two points in the same moduli space to count as a single vacuum. In this way, for instance, finiteness of vacua for e.g. M theory compactifications to five dimensions maps to the question of whether the Calabi-Yau threefold moduli space has a finite number of connected components (in fact, if ``Reid's fantasy'' is true \cite{Reid}, it would have only one).

Secondly, even in theories where a potential is generated, one may have infinite families of vacua, like $AdS_5\times S^5$ compactifications of IIB, which are indexed by an integer which can take any value. The point for these is that the quantum gravity cutoff $\Lambda$ goes to zero as we increase $N$, so if we fix the value of $\Lambda$, only a finite number of vacua area is allowed. More generally, according to the Distance Conjecture, the EFT cutoff goes to zero at the infinite distance limits, so that there is only a finite number of vacua (consistent EFTs) below any given finite cutoff.  In particular, it predicts \cite{Lust:2019zwm} that all the AdS vacua have extra space whose size scales with a power of $\Lambda$ and as $\Lambda \rightarrow 0$ the theory decompactifies to a higher dimensional theory. Therefore, the notion of the finiteness of vacua that we will adopt in this paper is as follows,\\

\emph{Finiteness of quantum gravity vacua: The number of low energy EFTs (after quotienting by moduli spaces) consistent with quantum gravity that are valid (at least) up to a fixed finite energy cutoff is finite.  }\\

The same type of argument given for the Distance Conjecture in Section \ref{sec:compact} -- based on the finiteness of the partition function in low enough dimension -- can be used here to argue for this conjecture\footnote{The above conjecture is also related to the notion that the volume of moduli space that can be described by a given EFT with cutoff is always finite, since infinite volume tails are always cut off by towers of states.}. If there were infinitely many vacua, the scalar field would probe all these vacua in quantum mechanics, implying in practice a non-compact configuration space and a divergent partition function.
If there were disconnected vacua (which is not inconsistent with the Cobordism Conjecture for a fixed cutoff) we would still be getting a divergent partition function for quantum gravity if we compactified to low enough dimension simply due to summing over the infinite choices.

This notion of the finiteness of vacua resonates with Cheeger's theorem \cite{cheeger}, highlighted in \cite{Acharya:2006zw}. The theorem states that in a (potentially infinite) sequence of Riemannian manifolds with metrics
such that the sectional curvatures are all bounded from above, the volumes are bounded below, and the diameters are bounded above,
there can only be a finite number of diffeomorphism types. The bound on the curvatures is necessary to have an EFT description given by compactification on a manifold, while the bounds on volumes and diameters are required to cut off the infinite distance tails of the field space where towers of states become light. Hence, these mathematical conditions can be replaced by the physical requirement of having an EFT description with a finite cutoff.\footnote{
It also goes along the lines of the conjecture made in \cite{Acharya:2006zw}, namely that the number of 4d vacua with an upper bound on the vacuum
energy, an upper bound on the compactification volume, and a lower bound on the mass of the lightest Kaluza-Klein tower, is finite. Again, everything can be replaced by the requirement of having an EFT description with a finite cutoff; the infinite tower of states then remains above this cutoff which effectively implies the bounds on volumes, couplings, etc.}

It is interesting to notice the similarity of the above notion of the finiteness of vacua with the Distance Conjecture. According to the Distance Conjecture, moduli spaces are only non-compact if there are towers of states becoming light so the cutoff goes to zero. Therefore, the moduli space for a given EFT with a finite cutoff is actually compact, in the same way that we also expect the number of vacua to be finite, and both statements follow from the finiteness of quantum gravity amplitudes.  It would be interesting to exploit this analogy further and try to define a distance among different EFTs or different topologies. An interesting candidate could come from the Gromov-Hausdorff metric, since it is proven that the space of $d$-dimensional Riemannian manifolds with an upper bounded diameter  and  Ricci $\leq(d-1)k$ is precompact in this metric. We leave this for future work. 

The above notion of the finiteness of vacua also constrains the possible scalar potentials that can arise in quantum gravity. For instance, a periodic scalar potential can only be valid for a finite field range, since otherwise it would give rise to infinitely many vacua. Analogously, potentials with infinitely many vacua are only allowed if they imply an infinite distance variation of the scalar fields, so that the number of vacua becomes finite below any finite cutoff, due to the tower of states. It also nicely matches with the finiteness of self-dual flux vacua shown in \cite{Grimm:2020cda} in the context of Calabi-Yau compactifications.

More generally, the finiteness of vacua might arise from a more broad concept of the finiteness of quantum gravity amplitudes. 
In the next Section, we will explain in more detail what we mean with the finiteness of amplitudes.

\section{Conclusion: On the quest for a general principle from finiteness of amplitudes}
\label{sec:finiteness}

The basic goal of this paper is to find the underlying principle behind many of the Swampland constraints that have been proposed so far. We have seen in the previous Sections how familiar Swampland conjectures, such as the Distance Conjecture and WGC, can arise from considerations about black hole physics and application of entropy bounds or finiteness of partition functions. 

In this Section, we wish to speculate on a generalization of these considerations, which is a natural extension of the finiteness of the Bekenstein-Hawking entropy of black holes.

The basic observation throughout this paper is that in a consistent quantum theory, physical amplitudes such as overlaps between two quantum states, expectation values, etc. must attain finite values. In an EFT framework, one often allows for pathological amplitudes -- for instance, in the SM without the Higgs field, scattering amplitudes would violate unitarity \cite{Cornwall:1974km,LlewellynSmith:1973yud} --. But in a full, UV complete theory, physical observables are always finite. In the context of quantum gravity, the finiteness of amplitudes can help us distinguish consistent models from those which are not, becoming a Swampland constraint. 

To discuss what the finiteness of quantum gravity amplitudes means, we need to explain which quantum gravity amplitudes we are talking about in the first place. We only know how to define precise observables, and hence precise amplitudes, for theories of gravity in the presence of asymptotic boundaries. Examples are the conformal boundary of AdS or null infinity for Minkowski theories. In these situations, we have a well-defined notion of what a quantum gravity amplitude is; examples are S-matrix elements in flat space, or CFT correlators in the AdS case. Although it would be very interesting, we do not know how to meaningfully discuss amplitudes for quantum gravity in a compact space; and in fact, the Baby Universe Hypothesis \cite{McNamara:2020uza} suggests that one should expect a one-dimensional Hilbert space. Therefore, in this paper we shall restrict to the non-compact case. In this case, the boundary degrees of freedom comprise a dual system where gravity is weakly coupled, and there is a well-defined notion of energy -- the energy of the asymptotic scattering states in flat space S-matrix, CFT Hamiltonian for AdS, etc. 
A proposal for a more precise statement would be that in any theory of quantum gravity, quantum gravity amplitudes defined in this way are finite.

We now comment on the application of our statements to the interesting particular case of partition functions. In field theory, partition functions often have volume divergences in non-compact spacetime. Similarly, one might guess that thermal effects can also give a divergent partition function in gravity; but for quantum gravity in Minkowski space, there is no finite temperature ensemble due to the Jeans instability towards production of black holes \cite{Gross:1981mg}. Therefore, the finiteness hypothesis predicts that something like the Jeans instability should occur to prevent this counterexample. It would be interesting to study the implications of finiteness for similar instabilities that can also appear at finite volume, such as the ones dual to $\mathcal{N}=4$
 SYM at suffiently large chemical potential \cite{Hollowood:2008gp}.

Physical divergences should not be confused with divergences in amplitudes of an effective field theory, which may not be UV complete. For instance, the effective field theory computation (coming from topological string amplitudes \cite{Bershadsky:1993cx,Bershadsky:1993ta}) of certain higher-derivative terms at  the conifold point of a Calabi-Yau space is divergent, but this is only because one is missing the contribution of a massless $D3$ brane \cite{Strominger:1995cz}. Including these, one recovers the correct result that these amplitudes are finite (but diverge logarithmically in the deep IR).

Our statement is true in AdS/CFT, since by holographic duality, any AdS amplitude is exactly equal to a CFT amplitude, which are manifestly finite. But the power of the statement relies on its generality, and the fact that it should hold beyond AdS/CFT. If the finiteness conjecture is true, it may bring us closer to an explanation for the finiteness of the string theory landscape. For instance, consider compactification of 10-dimensional string theory to two dimensions on an arbitrary 8-manifold $X_8$. In two dimensions, the geometry itself can fluctuate at arbitrarily small energy cost, so the wavefunction is spread over all moduli fields, and even topological transitions between two different choices of $X_8$ are only take a finite amount of energy. This means that all possible choices of $X_8$  will have nonzero contributions to quantum gravity amplitudes; if they are all positive and bounded below, they could induce a divergence, leading to the conclusion that the number of allowed $X_8$'s is finite. 

The finiteness conjecture is an extension of a principle first established in \cite{Hamada:2021bbz}, which used finiteness of the probe brane moduli space. A localized probe brane cannot have infinitely many internal states, as this would conflict with Bekenstein bound. This means that the moduli space of a probe brane must be compact, which is in itself a Swampland constraint \cite{Hamada:2021bbz,Bedroya:2021fbu}.

\subsection*{Acknowledgements}
We thank Mike Douglas for useful discussions and comments.
We thank the Summer Program of the Simons Center for Geometry and Physics for kind hospitality during the early stages of this work. The work of YH is supported by JSPS Overseas Research Fellowships. The work of MM, CV is supported by a grant from the Simons Foundation (602883, CV) and by the NSF grant PHY-2013858, which also supported IV in the early stages of this work. In the later stages IV was supported by grant RYC2019-028512-I from the MCI (Spain), and also gratefully acknowledges hospitality from CMSA at Harvard.

\appendix
\section{Entropy trouble for dilatonic black holes\label{app:stab}}

In this Appendix we repeat the entropy argument of Section \ref{sec:blackhole} for the particular case of an exponentially decreasing gauge coupling, common in string compactifications. This may help illustrate some of the general points made in the main text.

Let us first briefly review the main features of electric black hole solutions of the theory in \eqref{action} with $g=e^{a\phi}$. We refer the reader to \cite{Garfinkle:1990qj,Draper:2019utz} for further details.
The metric for the non-extremal electric solution of charge $Q$ and mass $M$ is given by
\beq
\label{ansatz}
ds^2=-fdt^2+f^{-1}dr^2+r^2R^2d\Omega_2^2
\eeq
where 
\begin{equation}\label{fR}f=\left(1-\frac{r_+}{r}\right)\left(1-\frac{r_-}{r}\right)^{\frac{1-a^2}{1+a^2}},\quad R= \left(1-\frac{r_-}{r}\right)^{\frac{a^2}{1+a^2}},\end{equation}
and $r_-,r_+$ are the inner and outer horizons respectively, which can be determined in terms of $Q,M$ \cite{Garfinkle:1990qj}. The extremal limit corresponds to $r_h\equiv r_-=r_+$ which occurs when $r_h=(1+a^2)M=\sqrt{1+a^2}Q_{ext}e^{-a\phi_0}$ and implies
$f=R^{2/a^2}$.
The horizon area, $A=r_h^2 R(r_h)^2$, which is proportional to the classical contribution to the entropy, becomes parametrically small for nearly extremal solutions, and is exactly zero for extremal solutions since $R(r\rightarrow r_h)\rightarrow 0$.

The scalar and electric field behave in general as \cite{Garfinkle:1990qj,Draper:2019utz}
\begin{equation}
\phi=-\phi_0+\frac1a\log R
,\quad F_{tr}=\frac{e^{-2\phi_0}Q}{r^2}.
\label{scfield}
\end{equation}
so the maximum dilaton excursion from $\infty$ to the horizon $r_h$ 
 \beq\Delta\phi= \frac1{2a} \log\frac{A}{r_h^2}\ \eeq
 is parametrically large for nearly extremal black holes, and diverges for extremal solutions since they become of zero-size.

By looking at \eqref{scfield}, one can see that  the gradient of the scalar indeed increases as we approach the black hole horizon. We can only trust the EFT description if the gradients and curvatures are small enough, that is, if\beq
\vert d\phi\vert^2=f(r)(\partial\phi)^2=\frac{a^2}{1+a^2}\frac{e^{-2a\phi_0}Q^2}{r^2A(r)}\leq \Lambda^2
\label{grad}
\eeq
where we have used \eqref{fR} and \eqref{scfield} for the case of extremal solutions. Recall that $A(r)$ here is the effective horizon area, $A(r)=r^2 R^2(r)$.
We should think of $\Lambda$ as the scale where the four-dimensional local field theory description  breaks down, like e.g. the string scale or the KK scale. As we know, in general $\Lambda$ depends on $\phi$, and our challenge is to determine exactly how. 

When the cutoff $\Lambda\rightarrow \infty$ this is saturated at the horizon $r=r_h$. Since these black holes have horizon zero area, in this case we would count an infinite number of pointlike particles as discussed in Section \ref{sec:blackhole}.  However, suppose that  $\Lambda$ is finite.
Then, the EFT breakdown happens at some value $r_\star$ before reaching the horizon.
At that value $r_\star$, the effective area of the black hole is given by
\beq
A(r_\star)=r_\star^2\left(1-\frac{Q\sqrt{1+a^2}e^{-a\phi_0}}{r_\star}\right) \label{area}
\eeq
and \eqref{grad} gets saturated for
\beq
Q^2=\frac{1+a^2}{a^2}e^{2a\phi_0}r_\star^2A(r_\star) \Lambda^2\ .\label{saturation}
\eeq
By replacing this into \eqref{entropy} with $L^2=A(r_\star)$,\footnote{From \eqref{area} and \eqref{saturation}, this equation has a solution only if $\Lambda < \frac{a}{1+a^2}L^{-1}$ for $Q=Q_\text{max}$. If this is not satisfied, there are infinitely many small black holes with arbitrary charges.} we get
\beq
\Lambda\leq \frac{a}{\sqrt{1+a^2}}e^{-a\phi_0}R(r_\star)
\eeq
implying that the cutoff must decrease faster than $R=\sqrt{A}/r$. Interestingly, by using \eqref{scfield}, this is equivalent to an exponential drop-off of the cutoff in terms of the field distance
\beq
\label{cutoff}
\Lambda \leq M_p\frac{a}{\sqrt{1+a^2}}\exp(-a\phi(r_\star))=\frac{\sqrt{2}a}{\sqrt{1+a^2}}\,g(r_\star)\,M_p
\eeq
as dictated by the Distance Conjecture. Furthermore, the exponential rate is such that it can also be written as proportional to the gauge coupling, which is precisely the cutoff dictated by the magnetic version of the WGC, as more generally derived in Section \ref{sec:blackhole}.

\section{Stability of small black holes}
In this Appendix we study the stability of these small black holes and show that they are sufficiently long lived to count as different species, as required in \eqref{entropy}.
When counting the contribution of the small black holes to the entropy, we have assumed that we can count the black holes as different species, so each black hole of a given charge contributes at least one unit to the entropy. This is justified if the interactions are weak and the black holes are long lived. We argued that this should be the case as the gauge coupling is small (and goes to zero at the horizon), but here we directly compute the Schwinger pair production rate, to make sure that it is exponentially suppressed. The expression for the rate of change of the charge can be obtained by integrating the Schwinger pair-production rate in the outside geometry, as
\beq
\frac{dQ}{dt}= \int \Gamma dr\wedge dt\simeq \int_{r_H}^\infty \sqrt{g} (qE)^2e^{-m^2/qE} dr
\eeq
where the electric field in our case is given by
\beq
qE(r)=\frac{Qqg(r)^2}{A(r)}=\frac{Qqe^{-2\phi_0}}{r^2}
\eeq
The result of this integral (for constant $m$) reads
\beq
\frac{dQ}{dt}\simeq \frac{1}{4} q^2 Q e^{-3\phi_0} \left(\sqrt{2}  \frac{2m^2 Q}{q} E_1\left(\frac{2 m^2 Q}{q}\right)+\sqrt{2} e^{-\frac{2 m^2 Q}{q}}-4 \sqrt{\pi } m \sqrt{q Q}\, \text{erfc}\left(\sqrt{2} m \sqrt{\frac{Q}{q}}\right)\right)
\eeq
where the special function $xE_1(x)$ is upper bounded by $\approx 0.26$ around $x=1$ and highly suppresed for $x>1$. We can also identify the exponential suppression characteristic of the Schwinger effect for large charges. To be long lived we need that the discharge rate is smaller than its mass $\frac{dQ}{dt}<M\sim e^{-\phi_0}Q$; this is automatically true due to the exponential factors.

This computation makes sense as long as $m^2\gg qE$ and the pair-production rate is suppressed; we have checked that the value of the radius at which this happens is smaller than $r_*$, and so it lies in the region outside of the reach of the EFT. The contribution to the decay rate coming from the core region, where the fieldstrengths are large, cannot be captured with the semiclassical analysis above. We leave an analysis of this interesting point to future work.




\bibliographystyle{JHEP}
\bibliography{Bibliography}

\end{document}